\documentclass[%
reprint,
 amsmath,amssymb,
apl,
aip,
 floats,
]{revtex4-1}

\usepackage{amsmath}
\usepackage{graphicx}% Include figure files
\usepackage{dcolumn}% Align table columns on decimal point
\usepackage{bm}% bold math
\newcommand{\ltsimeq}{\raisebox{-0.6ex}{$\,\stackrel{\raisebox{-0.2ex}{$\textstyle <$}}{\sim}\,$}}

\begin{document}

\preprint{APS/123-QED}

\title{A Josephson junction defect spectrometer for measuring two-level systems}

\author{M. J. A. Stoutimore}
 \email{micah.stoutimore@lps.umd.edu}
\author{M. S. Khalil}%
\affiliation{%
Laboratory for Physical Sciences, College Park, Maryland 20740, USA
}%
\affiliation{%
Center for Nanophysics and Advanced Materials, Department of Physics, University of Maryland, College Park, Maryland 20742, USA
}%
\author{C. J. Lobb}%
\affiliation{%
Center for Nanophysics and Advanced Materials, Department of Physics, University of Maryland, College Park, Maryland 20742, USA
}%
\affiliation{%
Joint Quantum Institute, University of Maryland, College Park, Maryland 20742, USA
}%
\author{K. D. Osborn}%
\affiliation{%
Laboratory for Physical Sciences, College Park, Maryland 20740, USA
}%

\date{\today}% It is always \today, today,
             %  but any date may be explicitly specified

\begin{abstract}
We have fabricated and measured Josephson junction defect spectrometers (JJDSs), which are frequency-tunable, nearly-harmonic oscillators that probe strongly-coupled two-level systems (TLSs) in the barrier of a Josephson junction (JJ). The JJDSs accommodate a wide range of junction inductances, $L_J$, while maintaining a resonance frequency, $f_{0}$, in the range of 4-8 GHz. By applying a magnetic flux bias to tune $f_{0}$, we detect strongly-coupled TLSs in the junction barrier as splittings in the device spectrum. JJDSs fabricated with a via-style Al/thermal AlO$_\text{x}$/Al junction and measured at 30 mK with single-photon excitation levels show a density of TLSs in the range $\sigma_{TLS}h \approx 0.4\text{-}0.5\text{ /GHz }\mu\text{m}^{2}$, and a junction loss tangent of $\tan\delta_{J} \approx 2.9\times10^{-3}$.

%\begin{description}
%\item[Usage]
%Secondary publications and information retrieval purposes.
%\item[PACS numbers]
%May be entered using the \verb+\pacs{#1}+ command.
%\item[Structure]
%You may use the \texttt{description} environment to structure your abstract;
%use the optional argument of the \verb+\item+ command to give the category of each item. 
%\end{description}
\end{abstract}

%\pacs{Valid PACS appear here}% PACS, the Physics and Astronomy
                             % Classification Scheme.
%\keywords{Suggested keywords}%Use showkeys class option if keyword
                              %display desired
\maketitle

%\tableofcontents

% \section{\label{sec:Intro}Introduction}

Recently, extensive research on superconducting circuits in the single-photon limit has shown that two-level system (TLS) defects that couple strongly to electric fields are an important source of loss. These TLSs have been identified in the barrier of Josephson junctions (JJs)\cite{Martinis2005,Simmonds2004,Kim2008}, in bulk dielectrics\cite{Martinis2005,Paik2010} and at materials interfaces\cite{Wenner2011}. Progress to improve the quality of superconducting qubits and resonators has been made through materials improvements in bulk dielectrics used for wiring and in capacitors\cite{OConnell2008,Paik2010} and by reducing the participation of the defects at materials' surfaces\cite{Paik2011}. Similarly, improvements have occured through a reduction in defect participation in the JJ barrier by adding capacitance in parallel to decrease the electric-field energy stored in the JJ\cite{Steffen2006,Schreier2008,Steffen2010} and by shrinking the area of the junction to reduce the probability of encountering a defect at the operating frequency of the device\cite{Martinis2005}. In other studies, the JJ barrier dielectric has been altered in order to directly improve device performance\cite{Oh2006,Kline2009,Nakamura2011,Weides2011}.

In this letter, we present the Josephson junction defect spectrometer (JJDS), which is a frequency-tunable resonator for measuring the quality of JJs. Unlike superconducting qubits, the JJDS is designed to have a nearly-harmonic oscillator potential which is optimized for defect detection rather than to allow for two-state quantum operations. Where previous efforts have focused on qubit-qubit coupling\cite{Berkley2003}, qubit-cavity coupling\cite{Wallraff2004} or qubit-TLS coupling\cite{Simmonds2004}, the JJDS is a new scheme that explores cavity-TLS coupling and provides a junction test circuit designed specifically to characterize and improve JJ material quality.

A significant advantage of the JJDS is that the critical current, $I_{c}$, of the JJ can vary by a factor of approximately 5 for a single device design so that different barrier thicknesses and materials can be studied with the same circuit. Proper operation of the JJDS requires only that the Josephson inductance, $L_{J0}=\Phi_{0}/(2 \pi I_{c})$, is greater than the total linear inductance, $L$, in parallel with the JJ. For comparison, with the phase qubit it is necessary to have $L/L_{J0} \approx$ 3-4. Additionally, the JJDS offers a robust alternative to qubit circuits because it is measured in the continuous regime without the need for quantum state preparation, and for the inductively coupled JJDS, we can use the same waveguide to couple dc and ac fields to the device. Finally, junction fabrication for the JJDS is compatible with sputtering and angle evaporation techniques\cite{Osborn2007}. As a proof of principal, we will show that for thermally grown barriers within via-style junctions the JJDS produces similar results to previous measurements on a phase qubit.

% \section{Design/Theory}
\begin{figure}[t] % tag b==bottom, also h==here, h!==force exactly here, t==top, p==special page for floats
\includegraphics[trim = 0.0cm 0.0cm 0.0cm 0.0cm, clip, scale=0.7]{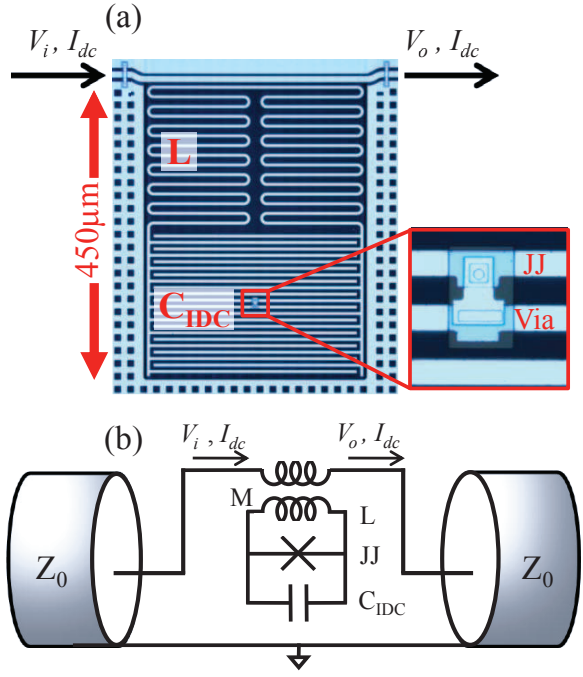} %trim left bottom right top
\caption{\label{fig:design} (a) An optical image of device 1 including a meandering wire inductor, $L$, and an interdigital capacitor, $C_{IDC}$, in parallel with a Josephson junction, $JJ$. (b) An effective circuit schematic of the physical device. The ac excitation, $V_{i}$, and magnetic flux bias, $\Phi_{ext} = MI_{dc}$, are applied through the transmission line, which is inductively coupled to the resonator.}
\end{figure}

We have designed and measured two versions of the JJDS. A microscope image of device 1 can be seen in Figure \ref{fig:design}a, while the design of device 2 is described elsewhere\cite{Osborn2007}. The effective circuit diagram for device 1 is shown in Figure \ref{fig:design}b. The circuit consists of a meandering inductor, $L$, in parallel with an interdigital capacitor (IDC), $C_{IDC}$. This capacitance is shorted by the JJ, which has a capacitance, $C_{J}$, and inductance, $L_{J}=\Phi_{0}/(2\pi I_{c} \cos\gamma)=L_{J0}/\cos\gamma$, where $\Phi_{0}$ is the magnetic flux quantum, $I_{c}$ is the junction critical current, $L_{J0}$ is the minimum inductance and $\gamma$ is the phase difference of the superconducting order parameter across the junction. Device 2 differs from device 1 primarily in that it is capacitively coupled to the transmission line and the magnetic flux bias is provided by a separate coil.

For both JJDS designs, $\gamma$ is proportional to the total flux through the inductor-junction loop such that $\gamma=2 \pi\Phi_{T}/\Phi_{0}$, and the circuit inductance is minimized at zero total flux. One can think of the JJDS as a capacitively-shunted rf-SQUID, which is made up of the same basic elements --– inductor, capacitor and JJ --– as the phase, flux and charge qubits. For the JJDS, one effect of the $C_{IDC}$ is to guarantee operation in the regime where the capacitive energy, $E_{c} =  e^{2}/(2C_{T})$, $C_{T} = C_{J}+C_{IDC}$ being the total capacitance, is less than the Josephson energy, $E_{J}=\frac{1}{L_{J0}}(\frac{\Phi_{0}}{2\pi})^{2}$. Because of this, we can treat the dynamics as that of a phase particle in a potential well given by:

\begin{equation} \label{Eq::NormPot}
u(\gamma) = \frac{1}{2}(\phi - \gamma)^2 - \frac{L}{L_{J0}}\cos\gamma,
\end{equation}
where $u = U\cdot L/(\Phi_{0}/2\pi)^2$ is the reduced potential energy and $\phi = 2\pi\Phi_{ext}/\Phi_{0}$ is the reduced external magnetic flux through the inductor-junction loop.

The JJDSs are fabricated in the limit of $L/L_{J0} \ltsimeq 1$ such that there is only one minimum in the potential well, and it has a nearly harmonic shape for a wide range of frequencies centered at $\phi = 0$. For device 1, the circuit parameters are: $C_{IDC} = 0.33$ pF, $L = 2.6$ nH and $I_{c} = 42.0$ nA, giving $L/L_{J0} \approx 0.3$, while for device 2, the circuit parameters are: $C_{IDC} = 1.68$ pF, $L = 400$ pH and $I_{c} = 0.9\text{ }\mu$A, giving $L/L_{J0} \approx 1.1$. In this single-well regime, we find that it is a reasonable approximation to treat the JJ as a variable linear inductor, $L_{J}(\phi)$, and calculate the resonance frequency using the harmonic value:

\begin{equation} \label{Eq::ResFreq}
f_{res} = \frac{1}{2\pi\sqrt{L_{T}C_{T}}} = \frac{1}{2\pi\sqrt{C_{J} + C_{IDC}}}\sqrt{\frac{L_{J} + L}{L_{J}L}}.
\end{equation}

To illustrate the validity of the linearized equation, we used the WKB method to quantize the energy levels of the well given by Equation \eqref{Eq::NormPot} with $L/L_{J0} = 1/3$. We find that the anharmonicity --– the difference in energy between the first and second levels of the well –-- is approximately $0.4\%$ for $0 \leq \Phi < 2\Phi_{0}/3$.

When measuring a JJDS, we apply an ac signal through an on-chip transmission line such that the average power coupled into the device is at the single-photon excitation level. In our current measurement setup, we mount the device in a dilution refrigerator with 20dB of attenuation at both the 1.5 K stage and the 30 mK stage to reduce noise on the input microwave signal. On the output microwave line, we have two circulators at the 30 mK stage and one at the 1.5 K stage to suppress thermal photons that propagate back to the sample. For device 1, the ac signal is combined through a bias tee with a dc signal at the 30 mK stage, which is then sent along the on-chip transmission line. Device 2 relies on a separate on-chip coil to couple dc signals to the JJDS. For both devices, the dc signal is used to set the magnetic flux bias through the inductor-junction loop of the JJDS. On resonance, the JJDS absorbs energy in the range of 4-8 GHz, which is limited only by the microwave electronic components in place. The transmitted signal is then amplified through a HEMT at 4 K.

% \section{Results}
\begin{figure}[b] % tag b==bottom, also h==here, h!==force exactly here, t==top, p==special page for floats
\includegraphics[trim = 0.1cm 0.2cm 0.2cm 0.1cm, clip, scale=0.48]{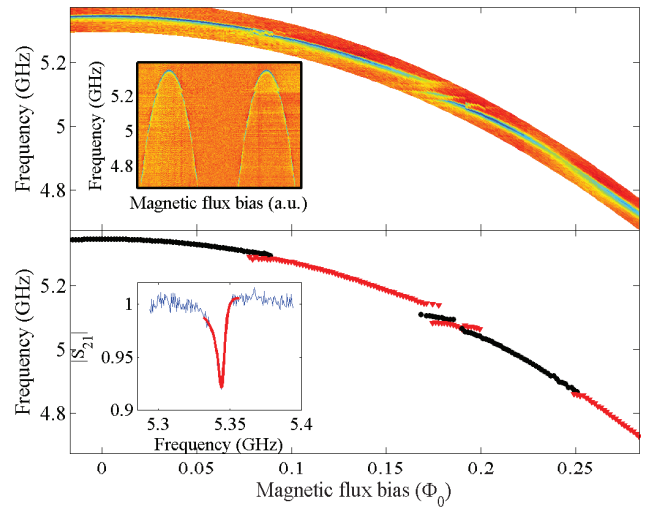}
\caption{\label{fig:spectrum} The spectrum on top shows the JJDS response to magnetic flux bias with splittings evident at several frequencies with the inset showing that the spectrum is periodic with applied magnetic flux as expected. The bottom panel shows the resonant frequency at each flux bias point determined by fitting the transmission curve as in the inset. The data is divided into alternating sections of $(\bullet)$ and $(\blacktriangledown)$ by the occurrence of splittings. Near one of these splittings, the curve is multi-valued and the splitting size is defined as the point of closest approach.}
\end{figure}

Figure \ref{fig:spectrum} shows a spectroscopic transmission measurement as a function of applied magnetic flux for device 1. We can see that as the applied flux modulates the junction inductance, the resonance frequency of the JJDS is also modulated. The top inset shows a wider magnetic flux bias sweep, where we observe the expected periodicity of the resonance frequency corresponding to the $\cos\gamma$ term in equation \eqref{Eq::NormPot}. Because the total energy in the device is below the single-photon limit, it remains in a nearly-coherent state and avoids bifurcation which has been observed previously in these devices at higher excitation levels and is useful for qubit readout\cite{Osborn2007}.

The inset in the bottom panel of Figure \ref{fig:spectrum} shows a slice of the spectroscopic image at $\Phi_{T} \approx 0$. The internal quality factor $Q_{i}$, which is a measure of the device quality, is extracted by fitting the transmission at each flux bias point using the diameter correction method\cite{Khalil2012}. For device 1, the internal quality factor is found to be $Q_{i} = 1/\tan\delta_{T} \approx 1.2\times10^{3}$, where $\tan\delta$ is the loss tangent for the device. 

From previous measurements\cite{Khalil2011}, we expect the internal quality factor of the IDC to be $Q_{i} = 10^{5}$. In the JJDS, the electric field energy is divided between the IDC and the junction based on their relative capacitances. Thus, the total loss is

\begin{equation} \label{Eq::TotalLoss}
\tan\delta_{T} = \frac{C_{J}\tan\delta_{J} + C_{IDC}\tan\delta_{IDC}}{C_{J} + C_{IDC}}.
\end{equation}

For the thermally grown barrier in a via-style junction shown in Figure \ref{fig:spectrum}, we find that $\tan\delta_{J} = 2.9\times10^{-3}$. This result is similar to previous measurements on phase qubit devices where the junction dominates the circuit capacitance\cite{Martinis2005}. In equation \eqref{Eq::TotalLoss}, we have assumed that the only sources of loss are the capacitive elements in the circuit. Although a full accounting of the loss in qubits is not yet understood, it is currently found that small JJ area qubits and large JJ area qubits with a small fraction of circuit capacitance in the JJ perform better than large JJ area qubits that are dominated by the JJ capacitance. To more directly evaluate the JJ, we look at the splittings that occur in the device spectrum.

The standard model for TLSs in glassy materials assumes a distribution that is inversely proportional to the tunneling matrix element\cite{Phillips1972}. When embedded in a JJ and allowed to interact with the electric field as in the JJDS, the distribution can be recalculated in terms of sample-dependent constants and the splitting size, $S$\cite{Martinis2005}. Integrating this distribution with respect to the splitting size shows that the expected number density is equal to $A_{J}\sigma_{TLS}\log(S)$ below the largest allowed splitting size, $S_{max}$. Here, $A_{J}$ is the area of the junction and $\sigma_{TLS}$ is the material dependent defect density that serves as a measure of the quality of dielectric material.

\begin{figure}[t]
\includegraphics[trim = 0.2cm 0.1cm 0.2cm 0.2cm, clip, scale=0.48]{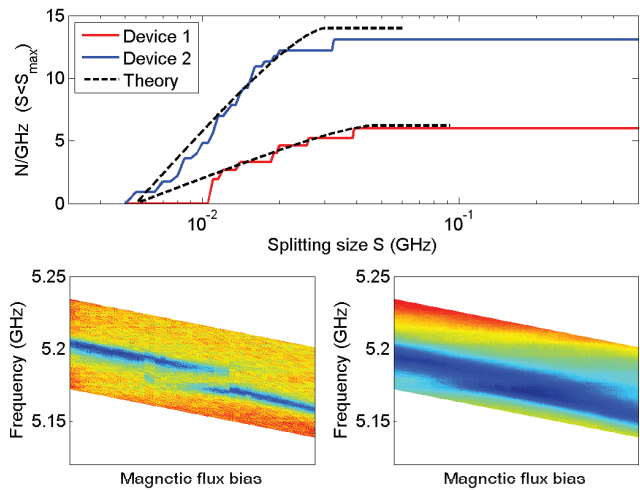}
\caption{\label{fig:integral} (a) By integrating the density of splittings versus the splitting size, we obtain the expected logarithmic dependence below the maximum splitting size. The maximum value is set by the total capacitance of the circuit. (b) A close-up of one splitting with excitation energy found from fitting to be $\bar{n}=0.002$ photons. (c) After increasing the input power 100x, the splitting can no longer be resolved.}
\end{figure}

Figure \ref{fig:integral}a shows the integral of splittings extracted from several spectra on samples using the device 1 and device 2 designs. The dashed lines represent fits of the data to the theoretical distribution. Device 1 is designed with an area, $A_{J}^{(1)}=3.1\text{ }\mu\text{m}^2$, while device 2 has an area $A_{J}^{(2)}=11.0 \text{ }\mu\text{m}^2$. From the slope of the fits, we can extract $\sigma_{TLS}^{(1)}h=0.5/\text{ GHz }\mu\text{m}^{2}$ and $\sigma_{TLS}^{(2)}h=0.4/\text{ GHz }\mu\text{m}^{2}$. These values are in agreement with previous measurements\cite{Martinis2005,Kline2009}.

The maximum splitting size is reached when the dipole moment of the TLS is parallel to the electric field in the JJ barrier. From our fits, we find that $S_{max}^{(1)}=46$ MHz and $S_{max}^{(2)}=30$ MHz. This value is expected to be proportional to $\sqrt{f_{res}/C_{T}}$, which accounts for the different values between the devices. In both cases, these values correspond to a ratio of the dipole moment to the barrier thickness of approximately 0.05, similar to the value of 0.06 found previously\cite{Martinis2005}.

To confirm that the splittings are indeed due to coupling of the JJDS circuit with TLS modes in the JJ dielectric, we can apply additional power to the circuit. For a TLS, it will continue to absorb power until it saturates, at which point it absorbs and emits energy at equal rates, effectively decoupling it from the JJDS mode so that the splitting disappears. In Figs. \ref{fig:integral}b and \ref{fig:integral}c, we show a splitting that is no longer resolvable when the input power is increased by 100 times. From fitting, we find that the average number of photons stored in the resonator is $\bar{n}=0.002$ and $\bar{n}=0.038$, respectively.

In conclusion, the JJDS can be used to create spectroscopic mappings of a JJ. From these measurements we can extract the areal density of TLSs, $\sigma_{TLS}$, and the loss tangent of the junction dielectric, $\tan\delta_{J}$, which we believe are the important figures of merit to compare the quality of junctions. We have shown that the JJDS gives similar values for these quantities to previous measurements on the phase qubit. Previous attempts to quantify loss in novel JJ barriers using the phase qubit architecture have used several different qubits on a single chip to ensure that a device with the required $L/L_{J}$ value is available\cite{Kline2009}. The flexibility of the JJDS to accommodate a wide range of junction parameters eliminates the need for this sort of sample design variation. Together with the ease of measurement of the JJDS, this makes it a useful alternative to qubits for studying junction quality.

K.D.O acknowledges helpful conversations with R. W. Simmonds. For M.S. and C.L., this research was supported by the Intelligence Advanced Research Projects Activity (IARPA) through the U.S. Army Research Office award No. W911NF-09-1-0351.

%\nocite{*}

\bibliography{JJDS}% Produces the bibliography via BibTeX.

\end{document}